\documentclass[conference]{IEEEtran}
\IEEEoverridecommandlockouts
\usepackage{cite}
\usepackage{bm}
\usepackage{multirow}
\usepackage{amsmath,amssymb,amsfonts}
\usepackage{algorithmic}
\usepackage{graphicx}
\usepackage{textcomp}
\usepackage{xcolor}
\usepackage{siunitx}
\usepackage[ruled,vlined]{algorithm2e}

\def\BibTeX{{\rm B\kern-.05em{\sc i\kern-.025em b}\kern-.08em
    T\kern-.1667em\lower.7ex\hbox{E}\kern-.125emX}}
\begin{document}

\title{Defect Detection by MIMO Wireless Sensing based on Weighted Low-Rank plus Sparse Recovery\\

\thanks{The work of U. S. K. P. M. Thanthrige and A. Sezgin is funded by the Deutsche Forschungsge-meinschaft (DFG, German Research Foundation)  Project--ID287022738 TRR 196 (S02) and the work of P. Jung is funded by the German Federal Ministry of Education and Research (BMBF) in the framework of the international future AI lab {``}AI4EO--Artificial Intelligence for Earth Observation: Reasoning, Uncertainties, Ethics and Beyond{''} (Grant number: 01DD20001). The two first authors have equal contributions.}
}

\author{\IEEEauthorblockN{Udaya S.K.P. Miriya Thanthrige$^{\star}$,~Ali Kariminezhad$^{\dagger}$,~Peter Jung$^{\ddagger,\ast}$ and Aydin Sezgin$^{\star}$} 
	\IEEEauthorblockA{\textit{$^{\star}$ Institute of Digital Communication Systems (DCS), Ruhr-University Bochum, Bochum, Germany.} \\
		\textit{$^{\dagger}$ Department of Autonomous Driving, Elektronische Fahrwerksysteme GmbH, Gaimersheim, Germany.}	\\}
	\textit{$^{\ddagger}$ Institute of Communications and Information Theory,
		Technical University Berlin, Berlin, Germany.}\\
	\textit{$^{\ast}$ Data Science in Earth Observation, Technical University of Munich, 82024 Taufkirchen/Ottobrunn, Germany.}\\
	\small{ \{udaya.miriyathanthrige,~aydin.sezgin\}@rub.de,~ali.kariminezhad@efs-auto.com,~peter.jung@tu-berlin.de.}
}
\IEEEpubid{\begin{minipage}{\textwidth}\ \centering \textbf{“This work has been submitted to the IEEE for possible publication. Copyright may be transferred without notice, after which this version may no longer be accessible.”.}\end{minipage}
}
\maketitle

\begin{abstract}
We present a compressive sensing based defect detection by multiple input multiple output (MIMO) wireless radar. Here, defects are inside a layered material structure, therefore, due to reflections from the surface of the layered material structure the defect detection is challenging. By utilizing a low-rank nature of the reflections of the layered material structure and sparse nature of the defects, we propose a method based on rank minimization and sparse recovery. To improve the accuracy in the recovery of low-rank and sparse components, we propose a non-convex approach based on the iteratively reweighted nuclear norm and iteratively reweighted $\ell_1$--norm algorithm. Our numerical results show that the proposed method is able to demix and recover the signalling responses of the defects and layered structure successfully from substantially reduced number of observations. Further, the proposed approach outperforms  the state-of-the-art clutter reduction approaches.
\end{abstract}


\section{Introduction}
\label{sec:intro}
Defect/object detection is important in many areas such as remote sensing, security, health, product quality assurance and many more~\cite{stoik2008nondestructive,yoon2009spatial,8903033,baker2007detection}. The main challenge of the detection of defects/objects is the strong clutter due to the reflection from the surface of the layered material structure \cite{tivive2011svd, tivive2014subspace}. The conventional clutter suppression methods such as spatial filtering (SF) \cite{yoon2009spatial}, background subtraction (BS) and subspace projection (SP) \cite{khan2010background} are
not effective in suppressing this kind of clutter as describe below. Reference data of the scene is required in BS. However, reference data may not be available in many scenarios. In the SP, it is difficult to determine the perfect threshold for clutter removal.\\ Many compressed sensing (CS)-based methods are proposed for object identification from very few measurements \cite{tang2014enhanced,tang2016indoor,7944209,tang2016radar,huang2009uwb,yoon2010through}. Moreover, joint clutter reduction and object recovery have been introduced recently in CS by considering that the clutter response has a low-rank structure while the response of the objects is sparse \cite{tang2016indoor, 7944209, tang2016radar}. However, finding matrices of lowest rank and sparsest vectors from compressed observations are NP-hard problems. Thus, it is common to consider the $\ell_1$--norm (absolute sum of elements) and the nuclear norm of a matrix (sum of singular values)~\cite{Recht2010} as convex relaxations to sparsity and rank. In both cases, non-convex approaches are known to perform better by providing tighter characterizations of rank and sparsity. In the $\ell_1$--minimization algorithms, larger coefficients are heavily penalized compared to the smaller coefficients and iteratively reweighted $\ell_1$--minimization has been considered in this context~\cite{candes2008enhancing, 5419071}. Similarly, nuclear norm minimization algorithms shrink singular values equally and tighter approaches based on reweighting has been considered~\cite{gu2017weighted,6858068, Fazel2003, Lu_2016}.\\
\IEEEpubidadjcol  \IEEEpubidadjcol 
In this work, we propose a novel iterative alternating direction method of multipliers (ADMM)-based algorithm for defect detection from few compressive measurements which combines iteratively reweighted nuclear norm and $\ell_1$--minimization. In particular, we consider a double-reweighted approach, i.e., wrt. nuclear and $\ell_1$--norm. In addition to that, we shed light on the improvement for multiple inner ADMM loops. Most of the work in the literature~\cite{wang2018robust, gu2017weighted} focus on low-rank plus sparse recovery (also known as robust principal component analysis (RPCA)) without considering the compressive sensing case~\cite{candes2011robust} and/or possibly with single reweighting.
In addition to that, we compare the performance of our method with state-of-the-art
clutter-reduction methods namely SF, SP and the low-rank plus sparse recovery method based on $\ell_1$--norm and nuclear norm minimization \cite{tang2016indoor}. The paper is organized as follows. We introduce the system model in Section \ref{sec:System_Model} and present the recovery algorithm in Section \ref{sec:LowRank_Sparse}. In Section \ref{sec:results}, we evaluate our approach numerically and finally Section \ref{sec:Con} concludes the paper. 

\section{System Model}
\label{sec:System_Model}
\begin{figure}[!t]
	\centering
	\includegraphics[width=1\linewidth]{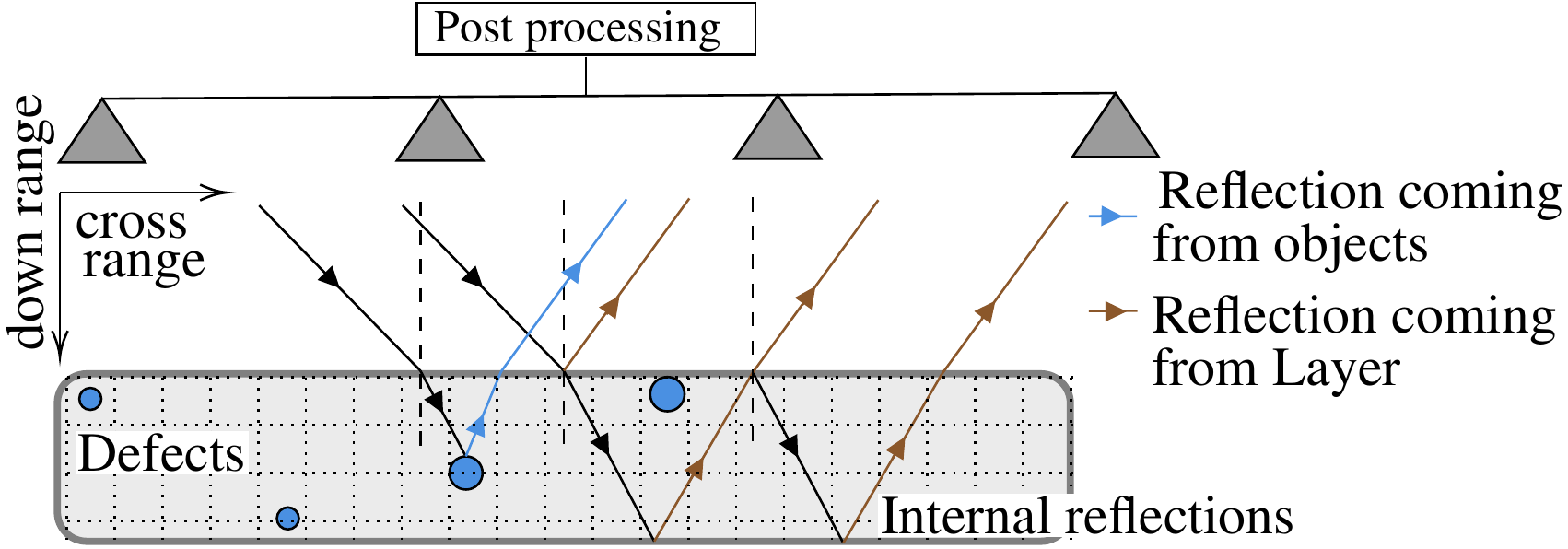}
	\caption{SFCW radar setup with $M$ transceivers which is used to identify defects.}
	\label{fig:fig1}
\end{figure}

In this work, a mono-static stepped-frequency continuous (SFCW) radar based setup is considered for defect detection as shown in Fig.~\ref{fig:fig1}. The SFCW radar system consists of $M$ transceivers parallel to the single-layered material structure. Also, equal amount of spacing is maintained between transceivers. Here, we consider $P$ static defects, which are inside the layered material structure. Each transceiver transmits a stepped-frequency signal consists of $N$ frequencies. The bandwidth of a single frequency band is given by $B/N$ where $B$ is the total bandwidth in Hz. The received signal of the $m$-th antenna for $n$-th frequency band $f_{n}$ is denoted by $y_{m,n}$. It includes of two major components, namely, the reflection of the layered material structure, i.e., $l_{m,n}$ and the reflection of the defects, i.e., $d_{m,n}$ as is given below
\begin{equation} \label{eq:rx_mn}
y_{m,n}= l_{m,n}+ d_{m,n}+ z_{m,n}.
\end{equation}
In~\eqref{eq:rx_mn} $z_{m,n}$ is the additive Gaussian noise. Now, $l_{m,n} \in\mathbb{C}$ in base-band is given by
\begin{equation}
	l_{m,n}=\sum\limits_{g=1}^{G+1}\alpha_{g}~e^{-j2\pi f_{n}\tau_{m,g}}.
	\label{eq:SS_eq_rx_lay_vec}
\end{equation}
The complex signal coefficient and the propagation delay of the $g$-th reflection of the layered structure are given by $\alpha_{g} \in\mathbb{C}$ and $\tau_{m,g}$, respectively. It is assumed that there are $G$ number of internal reflections within the layered structure as shown in Fig.~\ref{fig:fig1}. Now, the received signal of the layered structure for all $N$ frequencies is given by $\bm{l}_{m}=[l_{m,1},..., l_{m,N}]^T \in\mathbb{C}^N$. The reflection of the defects, $ d_{m,n} \in\mathbb{C}$ in base-band is given by
\begin{equation}
	d_{m,n}=\sum\limits_{p=1}^{P}\alpha_{p}~e^{-j2\pi f_{n}\tau_{m, p}}.
	\label{eq:SS_eq_rx_obj_vec}
\end{equation}
The complex signal coefficient of the $p$-th defect and the round-travel time of the signal from the $m$-th antenna location to the $p$-th defect are given by $\alpha_{p} \in\mathbb{C}$ and $\tau_{m, p}$, respectively. To this end, the received signal of the defects for all $N$ frequencies by the $m$-th antenna is given by $\bm{d}_{m}=[d_{m,1},..., d_{m,N}]^T \in\mathbb{C}^N$.\\ Next, the scene is hypothetically partitioned into a rectangular grid of size $Q$ as shown in Fig.~\ref{fig:fig1} to form a two-dimensional image of the scene. We define a vector $\bm{s} \in\mathbb{C}^{Q}$ which consists of all the $\alpha_{p}$ values of the defects. Then, $\bm{d}_{m}$ can be written as
$\bm{d}_{m}=\bm{A}_{m}\bm{s}$. The matrix $\bm{A}_{m} \in \mathbb{C}^{N \times Q}$ is generated based on the time delays between the antennas and grid locations. Let, $\tau_{m,q}$ is the propagation time delay between $m$-th antenna to the $q$-th grid location and back. To this end, $\text{exp}(-j2\pi f_{n}\tau_{m,q})$ represents the $(n,q)$-th element of $\bm{A}_{m}$. Now, we define, $\bm{Y} \in\mathbb{C}^{M \times N}$ which consists the received signals for all $M$ antennas and $N$ frequency bands. Now, $\bm{Y}$ is given by
\begin{equation}
	\bm{Y}= \bm{L}+ \bm{D}+ \bm{Z}.
\end{equation} 
\noindent Here, $\bm{L}$, $\bm{D}$ and $\bm{Z}$ $\in\mathbb{C}^{M \times N}$ are the received signals of the layered structure, defects and noise, respectively. We define a vectorization operator $(\text{vec}(\cdot))$, which is used to convert a matrix to a vector by stacking the columns of the matrix. To this end, $\text{vec}(\bm{L})$ and $\text{vec}(\bm{D})$ are given by $ [(\bm{l}_{1})^T,..., (\bm{l}_{M})^T]^T$ and $[(\bm{d}_{1})^T,...,(\bm{d}_{M})^T]^T$, respectively. Further, $\text{vec}(\bm{D})=\bm{A}\bm{s}$, with $\bm{A}= [(\bm{A}_{1})^T,...,(\bm{A}_{M})^T]^T \in\mathbb{C}^{MN \times Q}$.
\\The Compressed sensing (CS) approach considers that only a subset of antennas and frequencies are available. Here, this selection is done by using a selection matrix $\bm{\Phi} \in\mathbb{R}^{K \times MN}$. In each row of the selection matrix has a single non-zero element of equal to one that represents the chosen frequency of a selected antenna. The reduced data vector $\bm{y}_{cs} \!\in\!\mathbb{C}^{K}$  with $K\ll MN$ is given by
\begin{equation}
\begin{aligned}
	\bm{y}_{cs}&=\bm{\Phi}\left(\text{vec}(\bm{Y})\right),\\&=\bm{\Phi}\left(\text{vec}(\bm{L}+ \bm{D}+ \bm{Z})\right) \\&=\bm{\Phi}\left(\text{vec}(\bm{L})+\bm{A}\bm{s}+\text{vec}(\bm{Z})\right).
	\end{aligned}
	\label{eq:SS_eq_rx_all_cs}
\end{equation}
Now, our focus is to recover $\bm{L}$ and $\bm{s}$ using the compressive measurement set $\bm{y}_{cs}$.

\section{Low-rank plus sparse recovery}
\label{sec:LowRank_Sparse}
The response of the layered structure at different antennas shows strong similarity, i.e., $\bm{L}$ is low-rank. Moreover, the number of defects which are inside the layered material structure is less than grid size $Q$. Therefore,
we consider that the vector $\bm{s}$ in \eqref{eq:SS_eq_rx_all_cs}
is sparse (few non-zero entries). Thus, the estimation of $\bm{L}$ and
$\bm{s}$ from $\bm{y}_{cs}$ is formulated as a low-rank plus sparse
recovery problem. However, rank and sparsity minimization problems are usually NP-hard and therefore convex relaxations based on the nuclear norm and $\ell_1$--norm are considered. The resulting convex problems are meanwhile well understood on the theoretical level, but are often empirically outperformed by several non-convex approaches. A well-known approach is based on iterative reweighting of $\ell_1$--norm \cite{candes2008enhancing, 5419071} and nuclear norm \cite{gu2017weighted,6858068, Fazel2003, Lu_2016}. Motivated from \cite{candes2008enhancing, 5419071,gu2017weighted,6858068, Fazel2003, Lu_2016}, our approach is based on iterative reweighting of $\ell_1$--norm and nuclear norm. In more detail, we consider a double-reweighted approach, i.e., wrt. nuclear and $\ell_1$--norm which is not yet well studied in the literature for the compressive case. To this end, the estimation of $\bm{L}$ and $\bm{s}$ from $\bm{y}_{cs}$ is formulated as
\begin{equation}
	\label{eq:lowrank_sparse_rwnn}
	\begin{aligned}
		\min_{\bm{L},  \ \bm{s}} &\quad \beta_L~\left\Vert \bm{W_1}\bm{L}\bm{W_2}\right\Vert_{\star} + \beta_S \left\Vert \bm{W}_s\bm{s} \right\Vert_{1}, \\
		& \quad \  \ \text{s.t.} \  \left\Vert \bm{y}_{cs}-\bm{\Phi}~\text{vec}(\bm{L})-\bm{\Phi}\bm{A}\bm{s} \right\Vert_2^2 \  \le \ \epsilon.
	\end{aligned}%
\end{equation}
\noindent Here, $\beta_L$ and $\beta_S$ are given positive regularization parameters and a small positive constant $\epsilon$ is the noise bound. The $\ell_1$--norm of a vector and nuclear norm of a matrix are represented by $\left\Vert\cdot\right\Vert_1$ and  $\left\Vert \cdot \right\Vert_{\star}$, respectively. Further, $\bm{W_1}\! \in\mathbb{C}^{M \times M},\bm{W_2} \in\mathbb{C}^{N \times N}$ are complex weight matrices and $\mathbf{W}_s\!\in\!\mathbb{R}^{Q\times Q}\!$ is a non-negative diagonal weight matrix. Due to its multi-objective nature, the problem in eq. \eqref{eq:lowrank_sparse_rwnn} is challenging and therefore the alternating direction method of multipliers (ADMM) is used \cite{gabay1976dual, 7889039}. We denote the signal component at the $t$-th ADMM iteration as $(\cdot)^t$
and $\bm{u} \in\mathbb{C}^{K} $ and $\rho >0$ are the auxiliary coupling variable and penalty factor related to the ADMM approach, respectively.
Now, $\bm{L}$ at the $(t+1)$-th iteration is the solution of the nuclear norm minimization for fixed weights, which can be formulated as a semi-definite program (SDP) problem \cite{Fazel2003}, \cite{fazel2001rank}
\begin{subequations}
	\begin{align}
		(\bm{L})^{t+1} &= \arg\min_{\bm{L},\mathbf{L}_0,\mathbf{R}}\frac{\beta_L}{2}\big(\textrm{Tr}((\mathbf{W}_1)^{t}\mathbf{L}_0)+\textrm{Tr}((\mathbf{W}_2)^{t}\mathbf{R})\big)+ \notag \\& \quad   \dfrac{\rho}{2} \bigg\| \bm{\Phi}\text{vec}(\bm{L})+\bm{\Phi}\bm{A}(\bm{s})^t-\bm{y}_{cs}+\dfrac{1}{\rho} (\bm{u})^t  \bigg\|_2^2, \tag{\ref{eq:lowrank_sparse_ADMM_L}} \\
		&\text{s.t.}\
		\begin{bmatrix}
			\mathbf{L}_0 & \bm{L}\\
			\bm{L}^H & \mathbf{R}
		\end{bmatrix}\succeq 0. \nonumber
	\end{align}
	\label{eq:lowrank_sparse_ADMM_L}%
\end{subequations}
Notice that, the matrices $\mathbf{L}_0=\mathbf{L}_0^H\in\mathbb{C}^{M\times M}$ and $\mathbf{R}=\mathbf{R}^H\in\mathbb{C}^{N\times N}$ are auxiliary variables. The matrices $(\mathbf{W}_1)^{t}\in\mathbb{C}^{M\times M}$ and $(\mathbf{W}_2)^{t}\in\mathbb{C}^{N\times N}$ are weight matrices, which are prone to optimization as well. For given $(\mathbf{W}_1)^{t}$ and $(\mathbf{W}_2)^{t}$ at $t$-th iteration which are positive semi-definite ($\mathbf{W}_1, \mathbf{W}_2\succeq 0$), this problem is  a convex optimization problem. Therefore, to update $\mathbf{W}_1$ and $\mathbf{W}_2$, the eigenvectors of the $\mathbf{L}_0$ and $\mathbf{R}$ in the previous iteration are used. Let the eigenvectors of $\mathbf{L}_0$ and $\mathbf{R}$ be $\hat{\boldsymbol{\lambda}}$ and $\tilde{\boldsymbol{\lambda}}$, respectively. Now, $\mathbf{L}_0$ and $\mathbf{R}$ are given by		$\mathbf{L}_0=\mathbf{U}\textrm{diag}(\hat{\boldsymbol{\lambda}}^{(t)})\mathbf{U}^{H}$ and
		$\mathbf{R}=\mathbf{V}\textrm{diag}(\tilde{\boldsymbol{\lambda}}^{(t)})\mathbf{V}^{H}$. Next, to update $\mathbf{W}_1$ and $\mathbf{W}_2$ for the $(t+1)$-th iteration a decay function $f(\cdot)$ is used
\begin{equation}
	\begin{aligned}
		\boldsymbol{\gamma}^{(t)}_L=f\left(\hat{\boldsymbol{\lambda}}^{(t)}\right) \quad \text{and} \quad \boldsymbol{\gamma}^{(t)}_R=f\left(\tilde{\boldsymbol{\lambda}}^{(t)}\right),\\
	\end{aligned}
\end{equation}
\begin{equation}
	\begin{aligned}
		(\mathbf{W}_1)^{t+1}&=\mathbf{U}\textrm{diag}\left(\boldsymbol{\gamma}^{(t)}_L\right)\mathbf{U}^{H},\\
		(\mathbf{W}_2)^{t+1}&=\mathbf{V}\textrm{diag}\left(\boldsymbol{\gamma}^{(t)}_R\right)\mathbf{V}^{H}.
	\end{aligned}
	\label{eq:lowrank_sparse_ADMM_W}
\end{equation}
\begin{algorithm}[!t]
	{\fontsize{9pt}{12} \selectfont 
		\SetInd{0.2em}{0.4em}
		\SetKwInput{KwInput}{Input}                
		\SetKwInput{KwOutput}{Output} 
		\DontPrintSemicolon
		\textbf{Input:} $\bm{y}_{cs}$, $\epsilon=10^{-8}$, maximum number of \\ outer and inner iterations ($T$ and $J$), $\bm{\Phi}$, $\bm{A}$.\\ 
		\textbf{Initialization:} $\rho=10^{-2}$,~$\rho_{o}=1.05$,~$\rho_{\text{m}}=10^{3}$,~$t=0$,\\$(\bm{L})^{0}=\text{vec}^{-1}(\bm{\Phi}^	\dagger\bm{y}_{cs})$,~$(\bm{s})^{0}=\bm{0}_{Q}$,~$(\bm{W}_s)^{1}=\textrm{diag}(\bm{1}_{Q})$,\\  $(\bm{u})^{0}=\bm{0}_{K}$,~$(\mathbf{W}_1)^{1}=\textrm{diag}(\bm{1}_{M})$,~$(\mathbf{W}_2)^{1}=\textrm{diag}(\bm{1}_{N})$. 
		
		\While{$\left\Vert \bm{\Phi}~\text{\normalfont{vec}}(\bm{L})+\bm{\Phi}\bm{A}\bm{s}-\bm{y}_{cs} \right\Vert^2_2 > \epsilon$ or $t < T $}
		{
			\textbf {ADMM step 1: Update $\bm{L}$}\\
			\For { $j \gets1$ \textbf{To} $J+1$ \textbf{By} $1$}{ 
				\lIf {$j=1$} {$(\mathbf{W}_1)^{j}=(\mathbf{W}_1)^{t}$, $(\mathbf{W}_2)^{j}=(\mathbf{W}_2)^{t}.$} 
				Update $(\bm{L})^{j+1}$ by eq. \eqref{eq:lowrank_sparse_ADMM_L}.\\	
				Update $(\mathbf{W}_1)^{j+1}$ and $(\mathbf{W}_2)^{j+1}$ by eq. \eqref{eq:lowrank_sparse_ADMM_W} and \eqref{eq:WeightLR}.
			}
			$(\bm{L})^{t+1}=(\bm{L})^{j+1}$, $(\mathbf{W}_1)^{t+1}=(\mathbf{W}_1)^{j+1}$ and \\ $(\mathbf{W}_2)^{t+1}=(\mathbf{W}_2)^{j+1}$.\\
			\textbf {ADMM step 2: Update $\bm{s}$}\\
			\For { $j \gets1$ \textbf{To} $J+1$ \textbf{By} $1$}{
				\lIf {$j=1$} {$(\mathbf{W}_s)^{j}=(\mathbf{W}_s)^{t}$.}
				Update $(\bm{s})^{j+1}$ and $(\mathbf{W}_s)^{j+1}$ by eq. \eqref{eq:lowrank_sparse_ADMM_S} and eq. \eqref{eq:WeightS}.\\					
			}
			$(\bm{s})^{t+1}=(\bm{s})^{j+1}$ and  $(\bm{W}_s)^{t+1}=(\bm{W}_s)^{j+1}.$\\
			\textbf {ADMM step 3: Update $\bm{u}$}, $\rho$ and $t$\\
			Update $(\bm{u})^{t+1}$ by eq. \eqref{eq:lowrank_sparse_ADMM_U}. \\ 
			$\rho = \text{min} (\rho_{o} \times \rho,\rho_{\text{m}}),~ \text{and}~t=t+1.$
			\vspace{1pt}
		}	
		\KwOutput{$\bm{L}$, $\bm{s}$.}}
	\caption{\small Low-rank-plus-sparse recovery algorithm}
	\label{alternating_algo_admm}
\end{algorithm}
\normalfont
The operator $\textrm{diag}(\cdot)$ takes a vector as an input and returns a square diagonal matrix in which the main diagonal contains the vector elements and zeros elsewhere. 
Appropriate selection of decay function $f(\cdot)$ plays an important role in the algorithm. An overview of many known nonconvex surrogates for $\ell_0$--norm and nuclear norm is given in \cite{Lu2016}. In this work, the log-determinant heuristic \cite{Fazel2003} is used as the decay function $f$. Now, $\boldsymbol{\gamma}^{(t)}_L$ and $\boldsymbol{\gamma}^{(t)}_R$  in \eqref{eq:lowrank_sparse_ADMM_W} are given by 
\begin{align}
	\boldsymbol{\gamma}^{(t)}_L=(\hat{\boldsymbol{\lambda}}^{(t)}+\delta\bm{1}_M)^{-1}~\text{and}~
	\boldsymbol{\gamma}^{(t)}_R=(\tilde{\boldsymbol{\lambda}}^{(t)}+\delta\bm{1}_N)^{-1}.
	\label{eq:WeightLR}
\end{align}
Here, $\delta$ is a positive constant and $\bm{1}_B$  represents a vector of length $B$ with all elements equal to $1$. Next, $\bm{s}$ is updated as 
\begin{equation}
	\label{eq:lowrank_sparse_ADMM_S}
	\begin{aligned}
		(\bm{s})^{t+1} &= \arg\min_{\bm{s}} \  \beta_S \left\Vert (\mathbf{W}_s)^t\bm{s} \right\Vert_{1}+   \\ &   \dfrac{\rho}{2} \left\Vert \bm{\Phi}\text{vec}\left((\bm{L})^{t+1}\right)+\bm{\Phi}\bm{A}\bm{s} -\bm{y}_{cs}+\dfrac{(\bm{u})^t}{\rho}   \right\Vert_2^2.
	\end{aligned}
\end{equation}
The weight matrix for $t+1$-th iteration $(\bm{W}_s)^{t+1}$ is given by
\begin{equation}
	\label{eq:WeightS}
	(\bm{W}_s)^{t+1}=\textrm{diag}\left( \left(\left|(\bm{s})^{t+1}\right|+\delta\bm{1}_Q\right)^{-1}\right).
\end{equation}
Finally, $\bm{u}$ is updated as 
\begin{equation}
	\label{eq:lowrank_sparse_ADMM_U}
	(\bm{u})^{t+1}\!=\!(\bm{u})^{t}\!+\rho \!\left(\bm{\Phi}\text{vec}\!\left((\bm{L})^{t+1}\right) \!+\bm{\Phi}\bm{A}(\bm{s})^{t+1}\!-\bm{y}_{cs}\right).
\end{equation}
\\ The three main ADMM steps are given in eqs. \eqref{eq:lowrank_sparse_ADMM_L}, \eqref{eq:lowrank_sparse_ADMM_S} and \eqref{eq:lowrank_sparse_ADMM_U}, which are performed iteratively as shown in Alg.~\ref{alternating_algo_admm} to estimate the $\bm{L}$ and $\bm{s}$. In this work, we propose two versions of Alg.~\ref{alternating_algo_admm} named as low-rank plus sparse recovery with single inner ADMM loop ($\textrm{LRPSR}_\textrm{S}$) and low-rank plus sparse recovery with multiple inner ADMM loops ($\textrm{LRPSR}_\textrm{M}$), respectively. In the $\textrm{LRPSR}_\textrm{S}$, a single inner loop in the ADMM steps $1$ and $2$ is used (i.e., $J=0$  in Alg.~\ref{alternating_algo_admm}), while in the $\textrm{LRPSR}_\textrm{M}$ $J$ is set to $5$.
Note that, the multiple inner ADMM loops improve the performance compare to the singe inner ADMM loop. However, from our numerical results, we noticed that more than five ADMM inner loops do not significantly improve the performance. Therefore, we consider the number of inner ADMM loops as five ($J=5$). Further,  
$\text{vec}^{-1}(\cdot)$ and  $(\cdot)\dagger$ represent vector to matrix conversion and pseudo inverse, respectively. Notice that, in Alg.~\ref{alternating_algo_admm}, $\beta_L$, $\beta_s$ and $\delta$ are set to $1$, $1/\sqrt{\text{max}(M,N)}$ and $1$, respectively.
\section{Simulation results}
\label{sec:results}
First, a generic model was used to evaluate the performance of Alg.~\ref{alternating_algo_admm} generally. Next, simulations were performed based on the SFCW radar model given in Section \ref{sec:System_Model}.
\subsection{Tests with generic Gaussian model}
Here, the number of non-zero elements of the ground truth sparse vector $\bm{s}_{T}$ is set as $N_d=4$ and  the rank of the ground truth low-rank matrix $\bm{L}_T$ is set to $2$. The elements of $\bm{A}$ are generated as i.i.d.~$\mathcal{N}(0,1)$. $\bm{y}_{cs}$ is given by \eqref{eq:SS_eq_rx_all_cs} where the non-zero positions of $\bm{\Phi}$ are selected uniformly at random. 
The signal-to-noise ratio is $\text{SNR}:=\left\Vert\text{vec}(\bm{L}_T)+\bm{A}\bm{s}_{T}\right\Vert_2^2\big/\left\Vert\bm{Z}\right\Vert_F^2=20$dB and $M=10$, $N=20$ and $Q=256$. To quantify the recovery of defects, we use the target-to-clutter ratio (TCR). It is given by $	\text{TCR}:=\frac{Q-N_d}{N_d} \sum_{q\in A_{d}}\left\vert \bm{s}[q]\right\vert ^{2} \Big/ \sum_{q\notin A_{d}}\left\vert \bm{s}[q]\right\vert ^{2}$, where  $A_d=\{ i |~{s}_{T}[i] \neq 0 \}$ contains the actual locations of the $N_d$ defects. The average TCR[\si{\decibel}] for $100$ simulations is given in Table \ref{table:T1}.   
\begin{table}
\renewcommand{\arraystretch}{1.25}
	\caption{Average TCR (\si{\decibel}) for different
		clutter mitigation approaches (For SF and SP full data set is used ($100\%$)).}
	\resizebox{0.48\textwidth}{!}{
		\begin{tabular}{|c|l|l|l|l|l|}
			\hline
			Compression ratio ($K/MN$) & SP        & SF    & LRSR  & $\textrm{LRPSR}_\textrm{M}$ & $\textrm{LRPSR}_\textrm{S}$ \\ \hline
			$30\%$	&\multirow{3}{*}{$19.62$}   &\multirow{3}{*}{$47.00$}   &$21.44$   &$\bm{26.43}$   &$21.77$    \\ \cline{1-1} \cline{4-6}
			$40\%$	&   &   &$21.85$   &$\bm{31.29}$   &$26.26$    \\ \cline{1-1} \cline{4-6}
			$50\%$	&   &   &$22.07$   &$\bm{33.00}$   &$32.92$    \\ \hline
	\end{tabular}}
	\label{table:T1}
\end{table}
Here, the full data set, i.e., compression ratio $K/MN)=100\%$ is utilized for the spatial filtering (SF) and subspace projection (SP) and only for these methods it is assumed that $N_d$ is known. As such, SF only serves as a benchmark.
It can be observed that both the proposed approaches, $\textrm{LRPSR}_\textrm{M}$ and $\textrm{LRPSR}_\textrm{S}$, outperform SP and the low-rank plus sparse recovery method presented in \cite{tang2016indoor} (LRSR). Here, the performance improvement of our approaches ($\textrm{LRPSR}_\textrm{M}$ and $\textrm{LRPSR}_\textrm{S}$) over LRSR is mainly due to the iterative reweighing of  $\ell_1$--norm and nuclear norm minimization. Moreover, $\textrm{LRPSR}_\textrm{M}$ improves over $\textrm{LRPSR}_\textrm{S}$. This is due to the multiple inner ADMM loops.
\subsection{Tests with SFCW radar model}
Simulations were performed based on the SFCW radar model with a carrier frequency $f_c$ of $300$ GHz,
bandwidth $B$ as $5$~GHz, $N$ as $20$ and $M$ as $10$. The spacing between antenna elements set as half of the wavelength of $f_c$. Here, both height and length of the single-layered structure is $0.5$~m. The distance from the antenna array to the front surface of the layered structure is $1.5$~m. The radar scene is divided equally into a $16 \times 16$ grid. Moreover, we consider Rayleigh resolution of the radar to determine the grid size. The signal strength of the layered structure $\alpha_{g}$ is considered $10$ times stronger than the maximum signal strength of the defects $\alpha_{p}$. In the simulations, four defects are considered.
\begin{figure}[!t]
	\centering
	\includegraphics[width=0.95\linewidth]{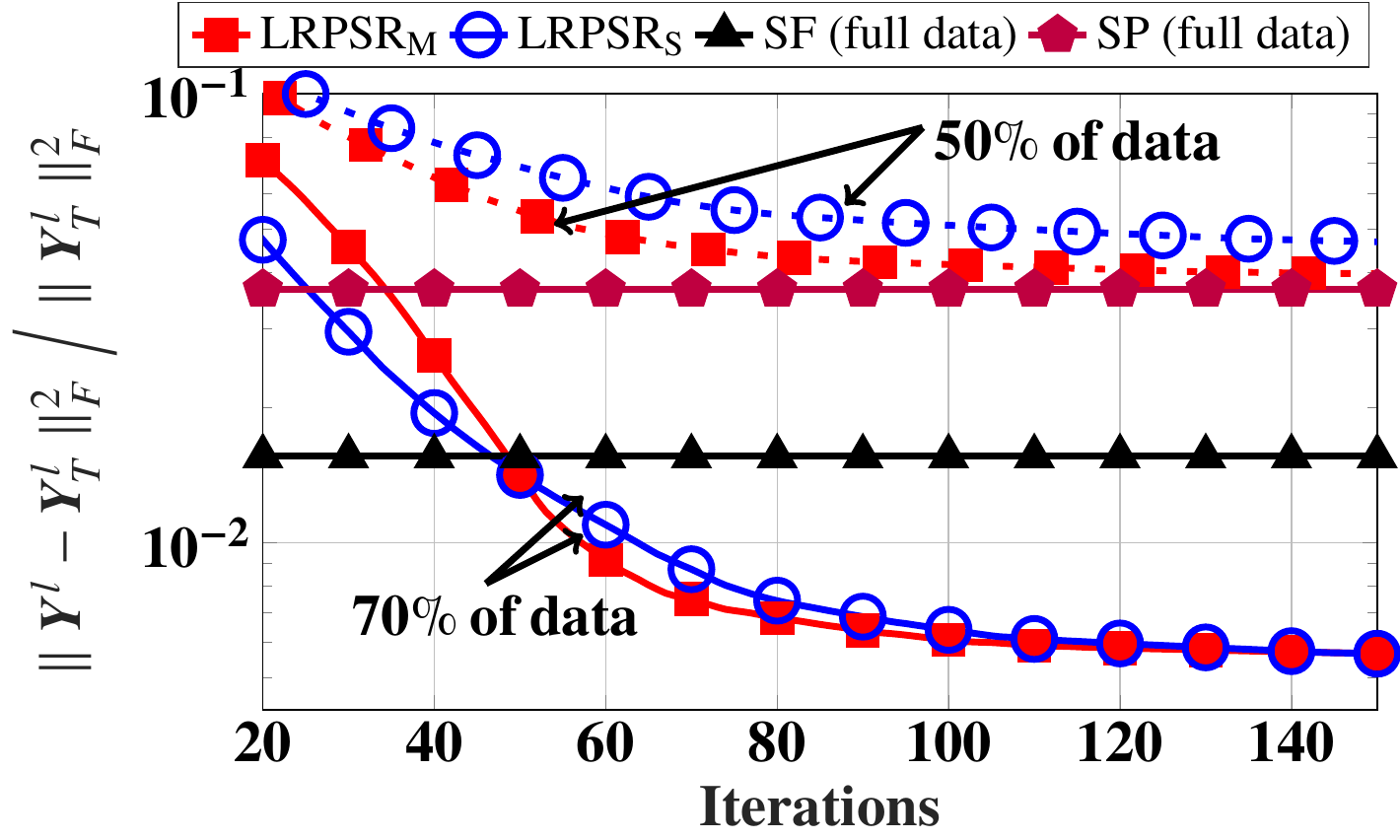}	
	\includegraphics[width=0.96\linewidth]{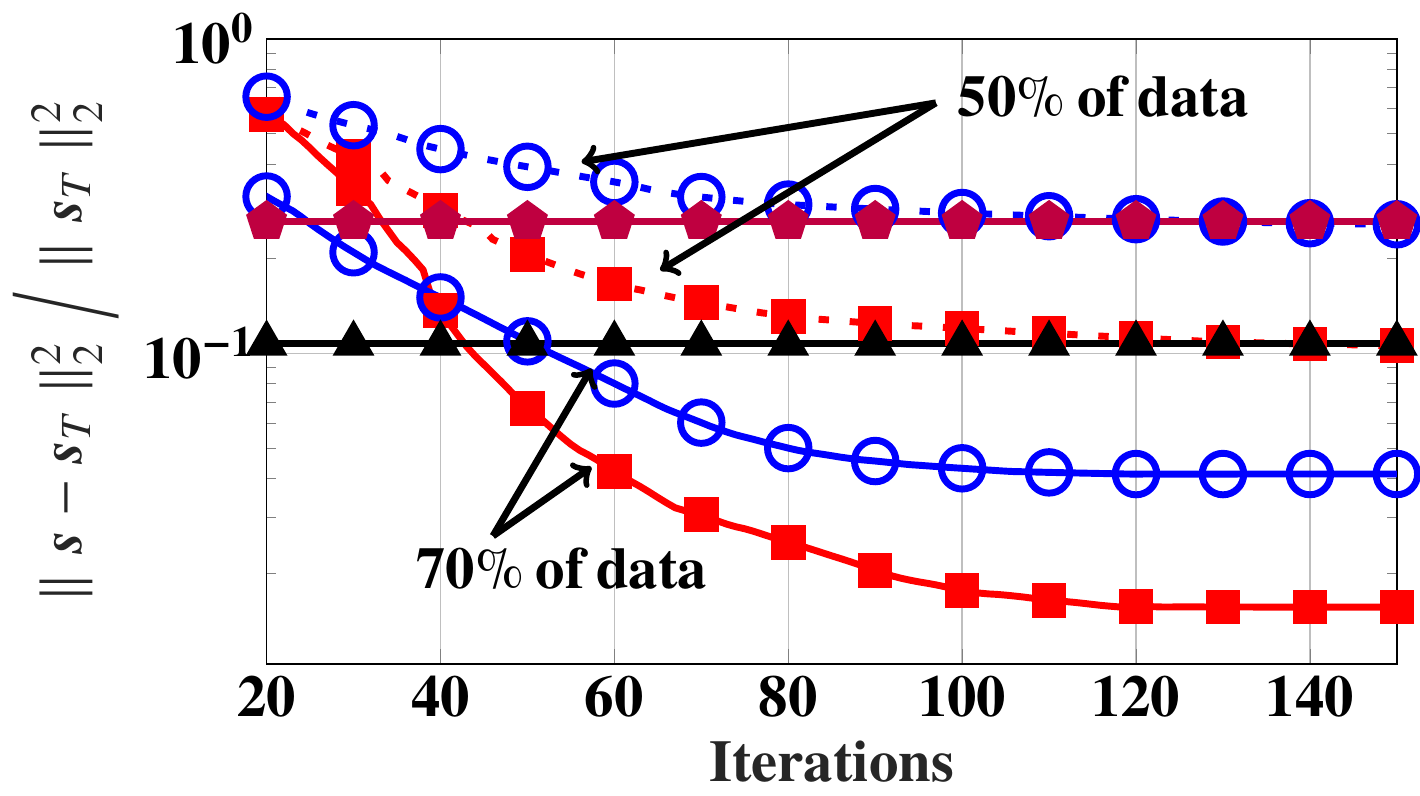}
	\caption{Average  normalized recovery error of low-rank (top) and sparsity (bottom) contributions.}
	\label{fig:lsrecovery}
\end{figure}
\begin{figure}[!t]
	\centering
	\hspace*{-1em}
	\includegraphics[width=0.98\linewidth]{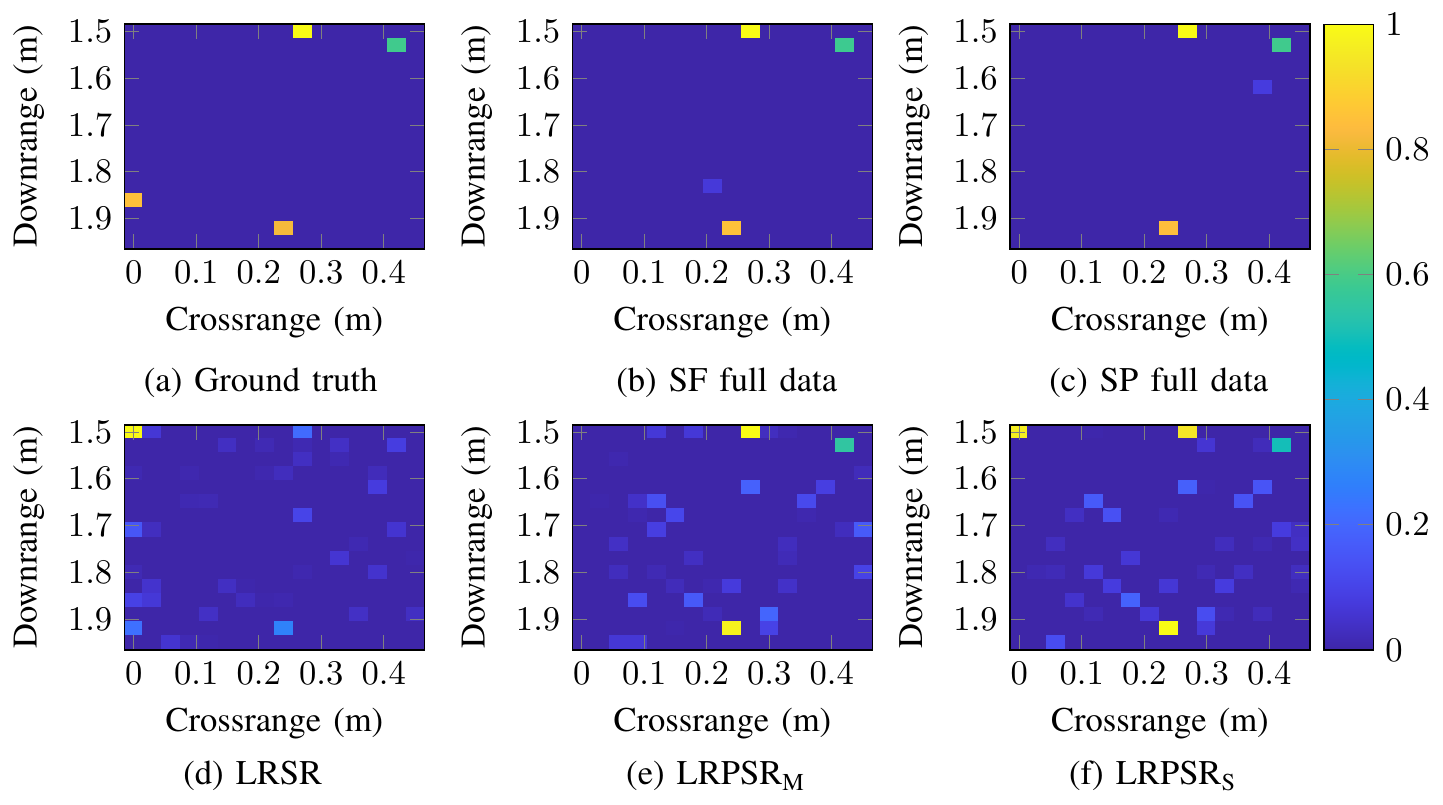}
	\caption{Image of the defects for SNR $20$ dB: (a) Ground truth. (b), (c) SF and SP with $100\%$ of data, (d), (e), (f) LRSR \cite{tang2016indoor}, $\textrm{LRPSR}_\textrm{M}$ and $\textrm{LRPSR}_\textrm{S}$ with $30\%$ of data ($K/MN=30\%$).}
	\label{fig:fig2}
\end{figure}
\\The images formed using different clutter-reduction methods are shown in Fig. \ref{fig:fig2}. Here, to obtain an image of the defects, the recovered vector $\bm{s}$ is reshaped into a matrix. Fig. \ref{fig:fig2}(a) shows the actual defect locations. Fig. \ref{fig:fig2}(b) and \ref{fig:fig2}(c) show the results of the SF and SP for the full data set. Fig. \ref{fig:fig2}(d), \ref{fig:fig2}(e) and \ref{fig:fig2}(f) show the result of the LRSR \cite{tang2016indoor} and the proposed approaches $\textrm{LRPSR}_\textrm{M}$ and $\textrm{LRPSR}_\textrm{S}$. It can be seen that the proposed $\textrm{LRPSR}_\textrm{M}$ and $\textrm{LRPSR}_\textrm{S}$ approaches are able to identify three out of four defects using only $30\%$ of the data. Further, it is observed that defect detection with $\textrm{LRPSR}_\textrm{M}$ and $\textrm{LRPSR}_\textrm{S}$ perform similar to the state-of-the-art SP and SF even with reduced data set. The TCR values of Fig. \ref{fig:fig2} for the SF, SP, LRSR, $\textrm{LRPSR}_\textrm{M}$ and $\textrm{LRPSR}_\textrm{S}$ are $71.4$ dB, $70.2$ dB, $30.8$ dB, $52.3$ dB, and $44.3$ dB, respectively. Higher TCR value means higher clutter reduction and the results show that the proposed approaches achieved satisfactory TCR values.\\
Fig. \ref{fig:lsrecovery} shows the average normalized mean square error (MSE) of the recovered low-rank and sparse components with respect to the iterations ($t$) of Alg.~\ref{alternating_algo_admm} for $20$ simulations. Here, SNR, $M$ and $N$ are set to $30$ dB, $10$, and $10$, respectively. The normalized MSE of the recovered low-rank and sparse components for a single simulation are given as $\left\Vert \bm{L}-\bm{L}_{T}\right\Vert_F^2 \big/ \left\Vert \bm{L}_{T}\right\Vert_F^2$
and $\left\Vert \bm{s}-\bm{s}_{T}\right\Vert_2^2 \big/ \left\Vert \bm{s}_{T}\right\Vert_2^2$, respectively. Based on Fig. \ref{fig:lsrecovery}, it can be observed that the $\textrm{LRPSR}_\textrm{M}$ outperforms $\textrm{LRPSR}_\textrm{S}$. Further, it can be observed that as the compression ratio decreases, the improvement of $\textrm{LRPSR}_\textrm{M}$ compared to the $\textrm{LRPSR}_\textrm{S}$ increases. Moreover, the proposed $\textrm{LRPSR}_\textrm{M}$ and $\textrm{LRPSR}_\textrm{S}$ outperform state-of-the-art SP and SF in low-rank and sparse recovery with only $70\%$ of data.
\section{Conclusion}
\label{sec:Con}
In this work, the low-rank plus sparse recovery approach is proposed for defect detection in the presence of strong clutter of the layered material structure. To this end, an iterative algorithm is developed based on ADMM for defect detection with iterative reweighed nuclear norm and $\ell_1$--norm minimization. Moreover, multiple inner loops in the ADMM low-rank and sparse update steps improve the low-rank and sparse recovery. The results show that the proposed approaches are able to improve defect detection with reduced data set. 

\bibliographystyle{IEEEbib}
\bibliography{refs}

\begin{thebibliography}{10}

\bibitem{stoik2008nondestructive}
C.D. Stoik, M.J. Bohn, and J.L. Blackshire,
\newblock ``Nondestructive evaluation of aircraft composites using transmissive
  {T}erahertz time domain spectroscopy,''
\newblock {\em Optics express}, vol. 16, no. 21, pp. 17039--17051, 2008.

\bibitem{yoon2009spatial}
Y.~S. Yoon and M.~G. Amin,
\newblock ``Spatial filtering for wall-clutter mitigation in through-the-wall
  radar imaging,''
\newblock {\em IEEE Trans. on Geosci. and Remote Sens.}, vol. 47, no. 9, pp.
  3192--3208, 2009.

\bibitem{8903033}
A.~{Kariminezhad} and A.~{Sezgin},
\newblock ``Spatio-temporal waveform design in active sensing systems with
  multilayer targets,''
\newblock {\em 2019 27th EUSIPCO}, pp. 1--5, 2019.

\bibitem{baker2007detection}
C~Baker, T~Lo, WR~Tribe, BE~Cole, MR~Hogbin, and MC~Kemp,
\newblock ``Detection of concealed explosives at a distance using terahertz
  technology,''
\newblock {\em Proceedings of the IEEE}, vol. 95, no. 8, pp. 1559--1565, 2007.

\bibitem{tivive2011svd}
F.~H.~C. Tivive, A.~Bouzerdoum, and M.~G. Amin,
\newblock ``An {SVD}-based approach for mitigating wall reflections in
  through-the-wall radar imaging,''
\newblock {\em 2011 IEEE Radar Conf.}, pp. 519--524, 2011.

\bibitem{tivive2014subspace}
F.~H.~C. Tivive, A.~Bouzerdoum, and M.~G. Amin,
\newblock ``A subspace projection approach for wall clutter mitigation in
  through-the-wall radar imaging,''
\newblock {\em IEEE Trans. on Geosci. and Remote Sens.}, vol. 53, no. 4, pp.
  2108--2122, 2014.

\bibitem{khan2010background}
U.~S. Khan and W.~Al-Nuaimy,
\newblock ``Background removal from {GPR} data using eigenvalues,''
\newblock {\em Proc. of the XIII Int. Conf. on Ground Penetrating Radar}, pp.
  1--5, 2010.

\bibitem{tang2014enhanced}
V.~H. Tang, A.~Bouzerdoum, S.~L. Phung, and F.~H.~C. Tivive,
\newblock ``Enhanced wall clutter mitigation for compressed through-the-wall
  radar imaging using joint bayesian sparse signal recovery,''
\newblock {\em 2014 IEEE ICASSP}, pp. 7804--7808, 2014.

\bibitem{tang2016indoor}
V.~H. Tang, A.~Bouzerdoum, S.~L. Phung, and F.~H.~C. Tivive,
\newblock ``Indoor scene reconstruction for through-the-wall radar imaging
  using low-rank and sparsity constraints,''
\newblock {\em 2016 IEEE Radar Conf.}, pp. 1--4, 2016.

\bibitem{7944209}
A.~{Bouzerdoum}, V.~H. {Tang}, and S.~L. {Phung},
\newblock ``A low-rank and jointly-sparse approach for multipolarization
  through-wall radar imaging,''
\newblock {\em 2017 IEEE Radar Conf.}, pp. 0263--0268, 2017.

\bibitem{tang2016radar}
V.~H. Tang, A.~Bouzerdoum, S.~L. Phung, and F.~H.~C. Tivive,
\newblock ``Radar imaging of stationary indoor targets using joint low-rank and
  sparsity constraints,''
\newblock {\em 2016 IEEE ICASSP}, pp. 1412--1416, 2016.

\bibitem{huang2009uwb}
Q.~Huang, L.~Qu, B.~Wu, and G.~Fang,
\newblock ``{UWB} through-wall imaging based on compressive sensing,''
\newblock {\em IEEE Trans. on Geosci. and Remote Sensing}, vol. 48, no. 3, pp.
  1408--1415, 2009.

\bibitem{yoon2010through}
Y.~S. Yoon and M.~G. Amin,
\newblock ``Through-the-wall radar imaging using compressive sensing along
  temporal frequency domain,''
\newblock {\em 2010 IEEE ICASSP}, pp. 2806--2809, 2010.

\bibitem{Recht2010}
B.~Recht, M.~Fazel, and P.~A. Parrilo,
\newblock ``{Guaranteed Minimum-Rank Solutions of Linear Matrix Equations via
  Nuclear Norm Minimization},''
\newblock {\em SIAM Review}, vol. 52, no. 3, pp. 471--501, 2010.

\bibitem{candes2008enhancing}
E.~J. Candes, M.~B. Wakin, and S.~P. Boyd,
\newblock ``Enhancing sparsity by reweighted $\ell_1$ minimization,''
\newblock {\em Journal of Fourier analysis and applications}, vol. 14, no. 5-6,
  pp. 877--905, 2008.

\bibitem{5419071}
D.~{Wipf} and S.~{Nagarajan},
\newblock ``Iterative reweighted $\ell_1$ and $\ell_2$ methods for finding
  sparse solutions,''
\newblock {\em IEEE J. Sel. Topics Signal Process}, vol. 4, no. 2, pp.
  317--329, 2010.

\bibitem{gu2017weighted}
S.~Gu, Q.~Xie, D.~Meng, W.~Zuo, X.~Feng, and L.~Zhang,
\newblock ``Weighted nuclear norm minimization and its applications to low
  level vision,''
\newblock {\em Int. journal of computer vision}, vol. 121, no. 2, pp. 183--208,
  2017.

\bibitem{6858068}
M.~{Malek-Mohammadi}, M.~{Babaie-Zadeh}, and M.~{Skoglund},
\newblock ``Iterative concave rank approximation for recovering low-rank
  matrices,''
\newblock {\em IEEE Trans. on Signal Proces.}, vol. 62, no. 20, pp. 5213--5226,
  2014.

\bibitem{Fazel2003}
M.~Fazel, H.~Hindi, and S.~P. Boyd,
\newblock ``{Log-det heuristic for matrix rank minimization with applications
  to Hankel and Euclidean distance matrices},''
\newblock {\em Proc. of the 2003 American Control Conf., 2003.}, vol. 3, pp.
  2156--2162 vol.3, 2003.

\bibitem{Lu_2016}
C.~Lu, J.~Tang, S.~Yan, and Z.~Lin,
\newblock ``Nonconvex nonsmooth low rank minimization via iteratively
  reweighted {Nuclear} norm,''
\newblock {\em IEEE Trans. Image Process.}, vol. 25, no. 2, pp. 829–839,
  2016.

\bibitem{wang2018robust}
S.~Wang, Y.~Wang, Y.~Chen, P.~Pan, Z.~Sun, and G.~He,
\newblock ``Robust {PCA} using matrix factorization for background/foreground
  separation,''
\newblock {\em IEEE Access}, vol. 6, pp. 18945--18953, 2018.

\bibitem{candes2011robust}
E.~J. Cand{\'e}s, X.~Li, Y.~Ma, and J.~Wright,
\newblock ``Robust principal component analysis?,''
\newblock {\em Journal of the ACM (JACM)}, vol. 58, no. 3, pp. 1--37, 2011.

\bibitem{gabay1976dual}
Daniel Gabay and Bertrand Mercier,
\newblock ``A dual algorithm for the solution of nonlinear variational problems
  via finite element approximation,''
\newblock {\em Computers \& mathematics with applications}, vol. 2, no. 1, pp.
  17--40, 1976.

\bibitem{7889039}
C.~{Lu}, J.~{Feng}, S.~{Yan}, and Z.~{Lin},
\newblock ``A unified alternating direction method of multipliers by
  majorization minimization,''
\newblock {\em IEEE Trans. Pattern Anal. Mach. Intell.}, vol. 40, no. 3, pp.
  527--541, 2018.

\bibitem{fazel2001rank}
M.~Fazel, H.~Hindi, S.~P. Boyd, et~al.,
\newblock ``A rank minimization heuristic with application to minimum order
  system approximation,''
\newblock {\em Proc. of the American control conf.}, vol. 6, pp. 4734--4739,
  2001.

\bibitem{Lu2016}
C.~Lu, J.~Tang, S.~Yan, and Z.~Lin,
\newblock ``{Nonconvex Nonsmooth Low Rank Minimization via Iteratively
  Reweighted Nuclear Norm},''
\newblock {\em IEEE Trans. on Image Proces.}, vol. 25, no. 2, pp. 829--839,
  2016.

\end{thebibliography}
\end{document}